\documentclass[12pt]{article}
\usepackage{graphics,epsfig, psfrag}
\def \HT{{\mathcal H}}
\def \LT{{\mathcal L}}

\def \HCT{\hat{\mathcal H}}
\begin{document}

\makeatletter
\@addtoreset{equation}{section}
\makeatother
\renewcommand{\theequation}{\thesection.\arabic{equation}}

\begin{titlepage}

\baselineskip =15.5pt
\pagestyle{plain}
\setcounter{page}{0}

\begin{flushright}
\end{flushright}

\vfil

\begin{center}
{\huge On Koopman-von Neumann Waves}
\end{center}

\vfil

\begin{center}
{\large D. Mauro}\footnote{e-mail: mauro@ts.infn.it}\\
\vspace {1mm}
Dipartimento di Fisica Teorica, Universit\`a di Trieste, \\
Strada Costiera 11, P.O.Box 586, Trieste, Italy \\ and INFN, Sezione 
di Trieste.\\
\vspace {1mm}
\vspace{3mm}
\end{center}

\vfil

\noindent 
In this paper we study the {\it classical Hilbert space} introduced by Koopman and von Neumann in their operatorial
formulation of classical mechanics. In particular we show that the states of this Hilbert space do not spread, differently
than what happens in quantum mechanics. The role of the phases associated to these classical "wave functions"
is analyzed in details. In this framework we also perform the analog of the two-slit interference experiment  
and compare it with the quantum case.
\vfil
\end{titlepage}
\newpage

\section{Introduction}

In their standard formulation classical and quantum mechanics are written in two completely different mathematical 
languages: for example  in classical mechanics observables are {\it functions} of a 2n-dimensional phase space, while
in quantum mechanics they are self-adjoint {\it operators} acting on an Hilbert space.  
In the literature
there are a lot of attempts to reformulate classical and quantum mechanics in similar forms, \cite{Wigner}-\cite{Moyal}. 
In this paper we shall
concentrate on the work of Koopman and von Neumann (KvN) who proposed, in 1931-32, an operatorial formulation of classical
mechanics, \cite{Koopman}-\cite{von Neumann}. The starting point of their work
is the possibility of defining an Hilbert space of {\it complex} and {\it square integrable} classical "wave"
functions $\psi(\varphi)$ such that $\rho(\varphi)\equiv|\psi(\varphi)|^2$ can be interpreted as a probability density
of finding a particle at the point $\varphi=(q,p)$ of the phase space. 
This $\rho$ has to evolve in time according to the well-known Liouville equation:
\begin{equation}
\displaystyle
i\frac{\partial}{\partial t}\rho(q,p)=\HCT\rho(q,p) \label{first}
\end{equation}
where $\HCT$ is the Liouville
operator $\HCT=-i\partial_pH\partial_q+i\partial_qH\partial_p$
and $H$ is the Hamiltonian of the standard phase space. In order to obtain (\ref{first}) Koopman and von Neumann postulated
the same evolution for $\psi$:
\begin{equation}
\displaystyle
i\frac{\partial}{\partial t}\psi(q,p)=\HCT\psi(q,p) \label{second}
\end{equation}
Since $\HCT$ contains only first order derivatives it is easy to check that 
from eq. (\ref{second}) one can
derive eq. (\ref{first}). This is something which does not happen in quantum mechanics where the analogue of 
$\HCT$ is the Schr\"odinger operator which contains second order derivatives. 

Now in quantum mechanics the complex character of the wave function is of fundamental importance for a lot of reasons:
while the modulus of the wave function gives the probability density $\rho$, the phase of $\psi$ brings in also
some physical information. In fact it is related to the mean value
of the momentum operator $\widehat{p}$ and it gives origin to the appearance of interference effects in two-slit experiments.
To answer the question if phases of classical "wave" functions $\psi(q,p)$ play an analogous role also in 
classical mechanics is the main goal of this paper which is organized as follows.

In section 2 we will briefly review the Koopman-von Neumann formalism and the associated 
functional formulation \cite{Gozzi2}.
In section 3 we will present a very simple but pedagogical example
that shows how, in the operatorial approach to classical mechanics, the lack of an uncertainty
relation between $\widehat{q}$ and $\widehat{p}$ and the different form of $\HCT$ with respect to the Schr\"odinger
operator
$\widehat{H}$ prevents the spreading of the wave functions. We shall also show 
that the phase and the modulus totally decouple in the equation   of motion of $\psi$ and that
the phases do not influence the expectation values of the observables of classical mechanics. 
In section 4, after spending
some words about the abstract Hilbert space of classical mechanics, we shall underline how a lot of the previous 
considerations
are a consequence of the {\it particular} representation we have chosen. It is the representation where the classical 
"wave" functions $\psi(q,p)$ are given by functions of $q$ and $p$. If we change
representation  the situation changes drastically. In particular if we use a representation where $p$ in $\psi(q,p)$
is replaced by a different variable $\lambda_p$, then phases begin to play a crucial role 
since they bring in physical information (e.g. the mean value of $\widehat{p}$).
In this representation $\psi(q,\lambda_p)$ and $\rho(q,\lambda_p)$ do not evolve in the same way, as 
$\psi(q,p)$ and $\rho(q,p)$ used to do, and there is, in a certain sense, a sort
of spreading of the wave function. Since in this representation phases seem to play a physical role 
we can say that it is 
necessary to consider, for mathematical consistency, 
an Hilbert space made up with {\it complex} and not with {\it real} wave 
functions. 

In quantum mechanics the complexity of the wave functions is one of the most important reasons for
the interference effects. Therefore a question that
arises quite naturally is: what happens in this operatorial approach to classical mechanics if we consider
complex wave functions? In section 5 we shall propose the classical analog of the two-slit experiment and we will
perform in details all the relevant calculations. We will show that the different form of the evolution operators
in the quantum and in the classical case leads to totally different results: the known interference phenomenon
at the quantum level and its non-appearance at the classical one. As we said above we tentatively link this difference
to the different form of the evolution operators but more work has to be done to get a deeper understanding
of the phenomenon. 

\section{Operatorial Approach to Classical Mechanics}

In quantum mechanics, starting from the Schr\"odinger equation for the wave function $\psi(x,t)$,
\begin{equation}
\displaystyle
i\hbar\frac{\partial\psi(x,t)}{\partial t}=\widehat{H}\psi(x,t)\;\;\Rightarrow\;\;i\hbar\frac{\partial\psi(x,t)}{\partial
t}=-\frac{\hbar^2}{2m}\frac{\partial^2}{\partial x^2}\psi(x,t) +V(x)\psi(x,t) \label{third}
\end{equation}
we have that the probability density $\rho(x,t)=|\psi(x,t)|^2$ satisfies a continuity equation of the form:
\begin{equation}
\displaystyle
\frac{\partial\rho}{\partial t}=-div\vec{j} \label{tre}
\end{equation}
where we have indicated with $j$ the probability density current:
\begin{equation}
\displaystyle 
\vec{j}=-\frac{i\hbar}{2m}\biggl(\psi^*\vec{\nabla}\psi
-\psi\vec{\nabla}\psi^*\biggr)
\end{equation}
Now if we write the wave function as $\psi=\sqrt{\rho}\,exp[iS/\hbar]$ we discover immediately that the phase $S$ enters
explicitly into the expression of the current probability density $\vec{j}$:
\begin{equation}
\vec{j}=\frac{\rho\vec{\nabla} S}{m} \label{cinque}
\end{equation}
As a consequence the equation of evolution of $\rho$ (\ref{tre}), couples the phase and the modulus square
$\rho$ of the wave functions, \cite{Sakurai}. We can also notice from eq. (\ref{tre})
that in quantum mechanics the probability density $\rho$ does not
evolve in time with the same Schr\"odinger Hamiltonian $\widehat{H}$ which gives the evolution of the wave function $\psi$.
 
The situation is completely different in the Hilbert space of classical mechanics. In fact, following Koopman and von
Neumann, we postulate  that the wave functions $\psi(\varphi,t)=\psi(q,p,t)$ evolve in time with the Liouvillian operator:
\begin{equation}
\HCT=-i\partial_{p_i}H\partial_{q_i}+i\partial_{q_i}H\partial_{p_i} \label{ht}
\end{equation}
according to the following equation:
\begin{equation}
\displaystyle
i\frac{\partial}{\partial t}\psi=\HCT\psi\;\;\Rightarrow\;\; \frac{\partial}{\partial t}\psi
=(-\partial_{p_i}H\partial_{q_i}+\partial_{q_i}H\partial_{p_i})\psi \label{lio1}
\end{equation}
We can think of (\ref{lio1}) as the analogue of the quantum Schr\"odinger equation, i.e. as the fundamental equation governing
the evolution of the vectors in the Hilbert space  of classical mechanics. These vectors are the complex wave functions
on the phase space obeying the normalizability condition 
$\int dqdp\,\psi^*(q,p)\psi(q,p)=1$. If we take
the complex conjugate of (\ref{lio1}) we obtain:
\begin{equation}
\displaystyle
\frac{\partial}{\partial t}\psi^*=(-\partial_{p_i}H\partial_{q_i}+\partial_{q_i}H\partial_{p_i})\psi^* \label{lio2}
\end{equation}
i.e. $\psi$ and $\psi^*$ satisfy the same equation.  Now if we multiply (\ref{lio1}) by $\psi^*$
and (\ref{lio2}) by $\psi$ and we sum the two equations we re-obtain eq. (\ref{first}), 
i.e. the evolution of $\rho(q,p)\equiv\psi^*(q,p)\psi(q,p)$
with the Liouvillian operator:
\begin{equation}
\displaystyle
\frac{\partial}{\partial t}\rho=(-\partial_{p_i}H\partial_{q_i}+\partial_{q_i}H\partial_{p_i})\rho
\;\;\Rightarrow\;\;  i\frac{\partial}{\partial t}\rho=\HCT\rho \label{lio3}
\end{equation}
So we have derived the standard Liouville equation (\ref{lio3}) as a 
consequence of having 
postulated eq. (\ref{lio1}). Moreover we notice that (\ref{lio3}) does not couple the modulus square
of $\psi$, i.e. $\rho$,
with the phase of $\psi$ differently from what happens in quantum mechanics, eq. (\ref{tre}) and (\ref{cinque}).

If we define the scalar
product between two wave functions $\psi$ and $\tau$ as $\langle\psi|\tau\rangle=\int d\varphi\,\psi^*(\varphi)\tau(\varphi)$ it is
easy to show that $\langle\psi|\HCT\tau\rangle=\langle\HCT\psi|\tau\rangle$, i.e. the Liouvillian $\HCT$ is a self-adjoint
operator; consequently the norm of the state $\langle\psi|\psi\rangle=\int d\varphi\,\psi^*(\varphi)\psi(\varphi)$ is conserved during
the evolution and we can consistently interpret $\rho(\varphi)=\psi^*(\varphi)\psi(\varphi)$ as a probability density in the phase space.

Before going on we want to show how the Liouvillian $\HCT$
 arises in a natural way in the functional approach to classical
mechanics described in ref. \cite{Gozzi2} which represents the path integral counterpart of the KvN
operatorial formulation. Let us ask ourselves: which is the {\it probability} of finding a particle at a point
$\varphi^a=(q^1,\dots,q^n;p_1,\dots,p_n)$ in phase space at time
$t$ if it was at $\varphi^a_i$ at the initial time
$t_i$? This probability is one if $\varphi^a_i$ and $\varphi^a$ are connected with a {\it classical path} $\phi^a_{cl}$,
i.e. a path that solves the classical equations of motion, and it is zero in all the other cases. So we can write:
\begin{equation}
P(\varphi^a,t|\varphi^a_i,t_i)=\widetilde{\delta}(\varphi^a-\phi^a_{cl}(t;\varphi_i)) \label{prob}
\end{equation}
where the RHS is a functional delta that forces us to stay on the classical path $\phi^a_{cl}$
associated with the initial condition $\varphi_i$. The functional delta can be rewritten as a delta 
on the Hamiltonian equations of motion $\dot{\varphi}^a=\omega^{ab}\partial_bH(\varphi)$ via the introduction 
of a suitable functional determinant:
\begin{equation}
\widetilde{\delta}(\varphi^a-\phi^a_{cl}(t;\varphi_i))=\widetilde{\delta}(\dot{\varphi}^a-\omega^{ab}\partial_bH)
det(\partial_t\delta^a_b-\omega^{ac}\partial_c\partial_bH) \label{Diracdelta}
\end{equation}
It can be shown, see eq. (3.51) of \cite{Gozzi2}, that the determinant in the previous equation is
formally independent of the fields $\varphi$ and so it can be put equal to one
if we are not interested in the study of nearby trajectories.  
Now if we exponentiate the Dirac delta in (\ref{Diracdelta}) via the introduction of auxiliary variables
$\lambda_a$ we get the following
path integral:
\begin{equation}
\displaystyle
P(\varphi,t|\varphi_i,t_i)=\int{\cal D}\varphi\,\widetilde{\delta}(\dot{\varphi}^a-\omega^{ab}\partial_{b}H)=\int{\cal D}\varphi{\cal
D}\lambda\; e^{i\int\LT \,dt} \label{cpi}
\end{equation}
where the Lagrangian $\LT$ is:
\begin{equation}
\LT=\lambda_a\dot{\varphi}^a-\HT
\end{equation}
with 
\begin{equation}
\HT=\lambda_a\omega^{ab}\partial_bH
\end{equation}
From the kinetic part of the Lagrangian we can deduce the form of the commutators \cite{Gozzi2} of the 
associated operatorial 
theory:
\begin{equation}
[\widehat{\varphi}^a,\widehat{\lambda}_b]=i\delta^a_b\;\;\Rightarrow\;\; 
[\widehat{q}^{\, i},\widehat{\lambda}_{q_j}]=[\widehat{p}^{\, i},\widehat{\lambda}_{p_j}]=i\delta^i_j\label{comm}
\end{equation}
The previous commutators\footnote{All the commutators different from (\ref{comm}) are identically
zero. In particular, differently from the quantum case, we have that $[q,p]=0$ which implies that 
we can determine with
an arbitrary precision the position and the momentum of a classical particle, like it happens in the standard {\it 
phase space} approach to classical mechanics.} can be realized considering $\varphi^a$ as multiplicative
operators and $\lambda_b$ as derivative ones:
\begin{equation}
\displaystyle
\widehat{\lambda}_b=-i\frac{\partial}{\partial\varphi^b}\;\;\Rightarrow\;\;
\widehat{\lambda}_{q_j}=-i\frac{\partial}{\partial q_j},\;\;
\widehat{\lambda}_{p_j}=-i\frac{\partial} {\partial p_j} \label{operatorial}
\end{equation}
From now on we will indicate this representation as the {\it Schr\"odinger representation} of classical
mechanics.
Using (\ref{operatorial}) $\HT$ becomes the following operator:
\begin{equation}
\HCT=-i\omega^{ab}\partial_bH\partial_a=-i\partial_{p_i}H\partial_{q_i}+i\partial_{q_i}H\partial_{p_i}
\end{equation}
that is exactly the Liouvillian $\HCT$ of eq. (\ref{ht}). This confirms that the path integral (\ref{cpi}) is the correct 
functional counterpart of the KvN operatorial theory. 

In rederiving (\ref{cpi}) we have started from the transition {\it probability} 
$P(\varphi,t|\varphi_i,t_i)$ of going from $\varphi_i$ to $\varphi$
in a time interval $t-t_i$. So we can say that the path integral (\ref{cpi}) gives a kernel of evolution for the classical
probability densities $\rho(\varphi,t)$, in the sense that if we know the probability density at the initial time $t_i$ we can
derive the probability density $\rho$ at any later time $t$ via the standard relation:
\begin{equation}
\rho(\varphi,t)=\int d\varphi_iP(\varphi,t|\varphi_i,t_i)\rho(\varphi_i,t_i)
\end{equation}
Let us remember that KvN postulated for $\psi$ the same equation of motion (\ref{lio1}) as for $\rho$. 
As a consequence their evolution can be represented as:
\begin{equation}
\psi(\varphi,t)=\int d\varphi_iK(\varphi,t|\varphi_i,t_i)\psi(\varphi_i,t_i)
\end{equation}
where the kernel of evolution $K(\varphi,t|\varphi_i,t_i)$ has the same expression
as the kernel of evolution $P(\varphi,t|\varphi_i,t_i)$ for the densities $\rho$.
The reader may be puzzled that the same kernel propagates both $\psi$ and $|\psi|^2$. We will see that
there is no contradiction in this by working out
in details the case of a free particle in appendix A. 

\section{Spreading and Phases of the Wave Functions}

One of the most characteristic effects of quantum mechanics is the spreading of the wave functions during their 
time evolution. Let us
consider a quantum system of a free particle in one dimension with Hamiltonian
$\displaystyle \widehat{H}=-\hbar^2\frac{\partial^2}{\partial x^2}$ and initially described by a gaussian wave function:
\begin{equation}
\displaystyle
\psi(x)=\frac{1}{\sqrt{\sqrt{\pi}a}}exp\biggl(-\frac{x^2}{2a^2}+\frac{i}{\hbar}p_ix\biggr) \label{initial}
\end{equation}
It is easy to check that at the beginning the mean value of the position $x$ is equal to zero and the uncertainty 
in the measurement
of the position is: $\displaystyle \overline{(\Delta x)^2}=\frac{a^2}{2}$. 
At time $t$ the wave
function will be:
\begin{equation}
\displaystyle
\psi(x,t)=N\cdot exp\biggl[-\frac{m}{2(ma^2+i\hbar
t)}\biggl(x-\frac{p_it}{m}\biggr)^2+\frac{i}{\hbar}\biggl(p_ix-\frac{p_i^2}{2m}t\biggr)\biggr] \label{ending}
\end{equation}
We can note how the coefficient $p_i$, which entered only into the phase
of the wave function at time $t=0$, managed to enter also the modulus of the wave function
at time $t>0$, see eq. (\ref{ending}). As a consequence the expectation value of the position $x$ at time $t$
depends explicitly on the original phase $p_i$;
in fact we have:
\begin{equation}
\overline{x(t)}=\int dx\,x\cdot |\psi(x,t)|^2=p_it/m
\end{equation}
So we see that the information about the mean value of $x$ is carried by terms appearing in the 
original phase of $\psi$. 
For the mean square deviation of $x$ we get:
\begin{equation}
\displaystyle
\overline{(\Delta x(t))^2}=\overline{(x-\bar{x}(t))^2}=\frac{a^2}{2}\biggl(1+\frac{t^2\hbar^2}{m^2a^4}\biggr) \label{aabb}
\end{equation}
So, at $t\to\infty$ the wave function will be totally delocalized:
\begin{equation}
\lim_{t\to\infty}\overline{(\Delta x(t))^2}=+\infty, \;\;\;\forall a>0
\end{equation}
This effect is present also if we prepare an initial state very sharply peaked around the origin,
in fact:
\begin{equation}
\displaystyle
\lim_{a\to 0}\overline{\Delta(x(t))^2}=\lim_{a\to 0}\biggl(\frac{a^2}{2}+\frac{t^2\hbar^2}{2m^2a^2}
\biggr)=+\infty, \;\;\; \forall t>0
\label{chiama}
\end{equation}
Note that the previous limit is $+\infty$ because of the presence of the parameter $a^2$ in the denominator. The relation
(\ref{chiama}) is not surprising: if $a\to 0$ at the beginning then we have a state perfectly localized in space,
i.e.
$\overline{(\Delta x)^2}\to 0$ and, from the Heisenberg uncertainty relations, we can deduce that
$\overline{(\Delta p)^2}\to +\infty$. So in this case the initial momentum is completely undetermined and, consequently,
even after an infinitesimal time interval, also the position of the particle becomes completely undetermined. These are the
well-known quantum mechanical effects. 

What happens in the operatorial version of classical mechanics?
As we have seen in the previous section, the evolution in time of the wave functions is generated by the Liouvillian
itself 
$\HCT=-i\partial_pH\partial_q+i\partial_qH\partial_p$
which, 
in the particular case of a one-dimensional free particle, has the following simplified form:
\begin{equation}
\HCT=
-i\frac{\widehat{p}}{m}\frac{\partial}{\partial q} \label{pmq}
\label{hfree}
\end{equation}
The free Liouvillian is essentially the product of two commutative operators: an operator of multiplication 
$\widehat{p}$ and an operator of derivation $\displaystyle -i\frac{\partial}{\partial q}$. So if we want
to diagonalize the $\HCT$ of eq. (\ref{pmq}) we have to diagonalize simultaneously both $\widehat{p}$
and $\displaystyle -i\frac{\partial}{\partial q}$. The eigenstates of $\widehat{p}$ associated to an arbitrary
real eigenvalue $p_0$ are the Dirac deltas $\delta(p-p_0)$; the eigenstates of 
$\displaystyle -i\frac{\partial}{\partial q}$ are instead the plane waves
$\displaystyle \frac{1}{\sqrt{2\pi}}exp[i\lambda_qq]$ and their correspondent eigenvalues\footnote{
We call $\lambda_q$ the eigenvalues of $\displaystyle -i\frac{\partial}{\partial q}$ since
$\displaystyle -i\frac{\partial}{\partial q}$ is just a representation of the abstract Hilbert space 
operator $\widehat{\lambda}_q$: see (\ref{operatorial}) and the next section.} are $\lambda_q$.
So the eigenstates of the free Liouvillian (\ref{pmq}) are just the product of the eigenstates of
$\widehat{p}$ and $\displaystyle -i\frac{\partial}{\partial q}$:
\begin{equation}
\tau_{\lambda_qp_0}(q,p)=\frac{1}{\sqrt{2\pi}}e^{i\lambda_qq}\delta(p-p_0) \label{eigenstates}
\end{equation}
and the associated eigenvalues are the product of the eigenvalues:
${\mathcal E}=\displaystyle \frac{\lambda_qp_0}{m}$. Now, suppose to take as initial wave function the
following double gaussian in $q$ and $p$:
\begin{equation}
\psi(q,p,t=0)=\frac{1}{\sqrt{\pi a b}}exp\biggl(-\frac{q^2}{2a^2}-\frac{(p-p_i)^2}{2b^2}\biggl) \label{double}
\end{equation}
where $a$ and $b$ are related to our initial uncertainty in the knowledge of $q$ and 
$p$:\break
$\overline{(\Delta q)^2}=a^2/2,\;\overline{(\Delta p)^2}=b^2/2$. Note that, 
since in classical mechanics $\widehat{q}$ and $\widehat{p}$
commute, there isn't any uncertainty relation. As a consequence $a$ and $b$ in (\ref{double})
are two completely independent paramaters and the product $\overline{(\Delta q)^2}\cdot\overline{(\Delta p)^2}$
can assume arbitrary small values. 

Now we can write the initial wave function (\ref{double}) 
as a superposition of the eigenstates (\ref{eigenstates}) of the Hamiltonian
$\HCT$ as:
\begin{equation}
\psi(q,p,t=0)=\int d\lambda_qdp_0\,c(\lambda_q,p_0)\tau_{\lambda_qp_0}(q,p)
\end{equation}
where the coefficients $c(\lambda_q,p_0)$ are:
\begin{eqnarray}
\displaystyle
c(\lambda_q,p_0)&=&\int
dqdp\,\tau^*_{\lambda_q,p_0}(q,p)\psi(q,p,t=0)=\nonumber\\
&=&\sqrt{\frac{a}{\pi b}}exp\biggl(-\frac{\lambda_q^2a^2}{2}-\frac{(p_0-p_i)^2}{2b^2}\biggr)
\end{eqnarray}
The wave function at $t$ is given by:
\begin{eqnarray}
\displaystyle
\psi(q,p,t)&=&\int d\lambda_qdp_0\,c(\lambda_q,p_0)\,exp[-i{\mathcal E} t]\,\tau_{\lambda_qp_0}(q,p)
\nonumber\\
&=&\frac{1}{\sqrt{\pi a b}}exp\biggl[-\frac{1}{2a^2}\biggl(q-\frac{p}{m}t\biggr)^2-\frac{(p-p_i)^2}{2b^2}\biggr]
\label{param}
\end{eqnarray}
Note that this $\psi(t)$ is related to $\psi(0)$ by the following equation:
\begin{equation}
\displaystyle
\psi(q,p,t)=\psi\biggl(q-\frac{pt}{m},p,0\biggr)=\psi\biggl(q-\frac{\partial H}{\partial p}t, p+
\frac{\partial H}{\partial q}t, 0\biggr) \label{psievolution}
\end{equation}
The previous relation could be inferred also from the path integral (\ref{cpi}). In fact in the case of a free particle,
see appendix A, the kernel
of propagation is correctly given by:
\begin{equation}
\displaystyle
K(\varphi,t|\varphi_i,0)=\delta\biggl(q-q_i-\frac{p_it}{m}\biggr)\delta(p-p_i) \label{kernel}
\end{equation}
from which we obtain immediately:
\begin{equation}
\displaystyle
\psi(\varphi,t)=\int d\varphi_i\,K(\varphi,t|\varphi_i,0)\psi(\varphi_i,0)=\psi\biggl(q-\frac{pt}{m},p,0\biggr)
\end{equation}
If we identify the modulus square of the wave function
$|\psi(q,p,t)|^2=\rho(q,p,t)$ with the probability density of finding a particle in a certain point of the phase space 
we have
that (\ref{psievolution}) implies:
\begin{equation}
\rho(q,p,t)=\rho\biggl(q-\frac{\partial H}{\partial p}t, p+\frac{\partial H}{\partial q}t, 0
\biggr)
\end{equation}
This is an equation in perfect agreement with the Liouville theorem $\displaystyle
\frac{d}{dt}\rho=0$. 

Let us now go back to eq. (\ref{param}) and calculate 
the mean values of the dynamical variables at a generic time $t$.
They are:
\begin{equation}
\displaystyle
\bar{q}=\int dqdp\, q\cdot|\psi(q,p,t)|^2=\frac{p_it}{m},\;\;\;\;\;\;\;\bar{p}=\int dqdp\,p\cdot|\psi(q,p,t)|^2=p_i
\label{equno}
\end{equation}
Note that, differently from quantum mechanics, the information on the mean values of $\widehat{p}$
is given by coefficients 
which appear in the modulus of the wave function.  
The mean square deviations are:
\begin{equation}
\overline{(\Delta q(t))^2}=\frac{a^2}{2}+\frac{b^2}{2}\frac{t^2}{m^2},\;\;\;\;\;\;\; 
\overline{(\Delta p(t))^2}=\frac{b^2}{2}
\label{var12}
\end{equation}
Therefore also in classical mechanics if $b\neq 0$ we have $\displaystyle \lim_{t\to\infty}
\overline{(\Delta q(t))^2}=+\infty$ and the wave function is totally delocalized. This is not strange if we consider
that we are giving a statistical description of a set of particles with a momentum  
distributed in a gaussian way around $p_i$. This means that we can have particles with momenta both greater
and smaller than $p_i$. These particles
cause a dispersion in the wave  function and, consequently, in the distribution of 
probability around the mean value of $q$.

If we instead consider the mechanics of a single particle we can measure exactly its position and momentum 
at the initial time. In this
case the terms that parametrize the gaussians (\ref{double}) go to zero ($a\to 0, b\to 0$) and the initial wave function
is a good approximation for the double Dirac delta $\delta(q)\delta(p-p_i)$. 
At the generic time $t$ the state will be described by another couple of Dirac deltas:
$\displaystyle \delta\biggl(q-\frac{p_it}{m}\biggr)\delta(p-p_i)$.
In this limiting case the variances are identically zero because, differently from the quantum case
(\ref{chiama}), 
in (\ref{var12}) the parameters $a^2$ and $b^2$ do not appear in the denominator:
\begin{equation}
\lim_{a,b\to 0}\overline{(\Delta q(t))^2}=\lim_{a,b\to 0}\biggl(\frac{a^2}{2}
+\frac{b^2}{2}\frac{t^2}{m^2}\biggr)=0,
\;\;\;\;\;\;\lim_{a,b\to 0}\overline{(\Delta p(t))^2}=\lim_{b\to 0}\frac{b^2}{2}=0\label{deltadelta}
\end{equation}
and the particle remains perfectly localized in the phase space at every instant of time $t$.

We feel that even this very simple and pedagogical example can be used to underline some very important differences between
the quantum and the classical operatorial approaches which are:\\ 
{\bf 1)} in classical mechanics we can prepare an initial wave function
that approximates a double Dirac delta in $q$ and $p$ since $\widehat{q}$ and 
$\widehat{p}$ are commuting operators and so there is no
uncertainty relation between them;\\
{\bf 2)} the classical dynamics given by $\HCT$ is such that, if we know with
absolute precision the position and the momentum at $t=0$, they remain perfectly determined at every instant of time
$t$ and there is not any spreading, see eq. (\ref{deltadelta});\\ 
{\bf 3)} the knowledge about the average momentum of the classical particle is 
brought by terms appearing in the modulus of the wave function. 

For a classical free particle it is easy to show that if we add a phase factor to an {\it arbitrary }
initial wave function this phase factor does not pass into the real part 
during the evolution differently from what happens in quantum mechanics, see eq. (\ref{ending}). This can be proved
as follows: every
initial classical wave function 
\begin{equation}
\psi(q,p)=F(q,p)exp[iG(q,p)] \label{effegi}
\end{equation}
can always be written as a superposition of the eigenstates of the free Liouvillian 
(\ref{pmq}) in the following way:
\begin{equation}
\psi(q,p)=\int d\lambda_qdp_0\,c(\lambda_q,p_0)\tau_{\lambda_qp_0}(q,p)
\end{equation}
where the eigenstates $\tau_{\lambda_q,p_0}(q,p)$ are given by eq. (\ref{eigenstates}). 
So the coefficients $c(\lambda_q,p_0)$ are basically the Fourier transform, in the
$q$ variable only, of $\psi(q,p_0)$:
\begin{equation}
\displaystyle
c(\lambda_q,p_0)=\frac{1}{\sqrt{2\pi}}\int dq \,e^{-i\lambda_q q}F(q,p_0)e^{iG(q,p_0)}
\end{equation}
At this point we can 
find the free evolution of the wave function in the usual way:
\begin{eqnarray}
\displaystyle
&&\psi(q,p,t)=\int d\lambda_qdp_0\,c(\lambda_q,p_0)\,exp[-i{\mathcal E} t]\,\tau_{\lambda_qp_0}(q,p)=\nonumber\\
&&=\int d\lambda_q \,c(\lambda_q,p)exp\biggl[i\lambda_q\biggl(q-\frac{pt}{m}\biggr)\biggr]=
F\biggl(q-\frac{pt}{m},p\biggr)exp\biggl[iG\biggl(q-\frac{pt}{m},p\biggr)\biggr] \label{psiqpt}
\end{eqnarray}
and we obtain a result that is again in perfect agreement with the kernel of evolution (\ref{kernel}).
Therefore, in the case of a free particle, we have that 
for {\it every} initial wave function $\psi(q,p,t=0)=F(q,p)exp[iG(q,p)]$
the probability density 
$|\psi(q,p,t)|^2$ does not depend on the phase $G(q,p)$ not only at the beginning, but also at any
later time; in fact from (\ref{psiqpt}) we have that $\displaystyle |\psi(q,p,t)|^2=F^2\biggl(
q-\frac{pt}{m},p\biggr)$. 
This has some consequences also on the expectation values of the observables. If we assume that observables
in classical mechanics are only the functions $O(\varphi)$ or, in operatorial terms, only the operators
$O(\widehat{\varphi})$ then it is easy to check that their
expectation values do not depend on the phase $G$ of the wave function (\ref{effegi}):
\begin{equation}
\langle O\rangle=\int d\varphi F^*(\varphi)exp[-iG(\varphi)]O(\varphi)F(\varphi)exp[iG(\varphi)]=\int d\varphi
F^*(\varphi)O(\varphi)F(\varphi)
\end{equation}
This is true also at later times $t$ because of the form (\ref{psiqpt}) of the wave function at $t>0$. 

This independence from the phase $G$ would not happen if the observables were dependent also on $\lambda$: 
${\mathcal O}(\widehat{\varphi},\widehat{\lambda})$, because $\widehat{\lambda}$ is a derivative operator and 
it would give
\begin{eqnarray}
\displaystyle
\langle {\mathcal O}\rangle &=& \int d\varphi\,F^*(\varphi) exp[-iG(\varphi)]{\mathcal O}
\biggl(\varphi,-i\frac{\partial}{\partial\varphi}\biggr)
F(\varphi) exp[iG(\varphi)] =\nonumber\\
&=& \int d\varphi \,F^*(\varphi) Q(\varphi, F, F^{\prime}, G, G^{\prime})F(\varphi)
\end{eqnarray}
So $\langle{\mathcal O}\rangle$ would be a function of $F(\varphi)$, its derivative $F^{\prime}$ with respect to $\varphi$,
but also of the phase $G(\varphi)$ and its derivative $G^{\prime}$.
These considerations can be
extended from the free particle case to a physical system characterized by a generic Hamiltonian $\HCT$. 
In fact the solution
of the equation:
\begin{equation}
\displaystyle
i\frac{\partial}{\partial t}\psi(q,p,t)=\HCT\psi(q,p,t)
\end{equation}
is given by \cite{Sudarshan}:
\begin{equation}
\psi(q,p,t)=\psi(\bar{q}(q,p,t),\bar{p}(q,p,t)) \label{trasc}
\end{equation}
where $\bar{q}$ and $\bar{p}$ are the solutions of the equations:
\begin{equation}
\displaystyle
\dot{\overline{q}}_j(q,p,t)=-\frac{\partial H(\bar{q},\bar{p})}{\partial\bar{p}_j},\;\;\;
\dot{\overline{p}}_j(q,p,t)=\frac{\partial H(\bar{q},\bar{p})}{\partial\bar{q}_j}
\end{equation}
with the initial conditions $\bar{q}_j(q,p,0)=q^0_j,\;\; \bar{p}_j(q,p,0)=p^0_j$. So,
according to eq. (\ref{trasc}), the evolution of a classical system via the Liouvillian does not modify,
in the Schr\"odinger representation, the functional form of $\psi$ provided we write it using 
the $\bar{q},\bar{p}$. This has, as an immediate consequence, that if we
take a wave function without any phase at the initial time, phases {\it cannot} be generated during the evolution:
\begin{equation}
\HCT:\;\;\;\psi(without\;\; phases \;\, t=0) \;\;\;\longrightarrow\;\;\; \psi(without \;\;phases \;\, t)
\end{equation}
In quantum mechanics instead, even if we start from a wave function that does not present phases, these ones
will be created in general at later times via the operator $\widehat{H}$:
\begin{equation}
\widehat{H}:\;\;\;\psi(without \;\;phases\;\, t=0) \;\;\;\longrightarrow \;\;\;\psi(with \;\, phases \; \, t)
\end{equation}
This can be seen for example from (\ref{initial}) and (\ref{ending}). Even if we start with no phase $p_i=0$,
at time $t$ we get that the wave function (\ref{ending}) becomes:
\begin{equation}
\displaystyle
\psi= N\cdot exp\biggl[-\frac{m}{2(ma^2+i\hbar t)}x^2\biggr]
\end{equation}
and it has a phase because of the term $i\hbar t$ in the denominator. All this is a consequence of the fact that the phase
and the modulus of a wave function interact in quantum mechanics as one can see from standard text books, \cite{Messiah}.
In fact writing
\begin{equation}
\psi(x)=A(x) exp\biggl[\frac{i}{\hbar}S(x)\biggr]
\end{equation}
and equating the real and imaginary part of the Schr\"odinger equation (\ref{third}) 
we obtain the following two equations for $A(x)$ and $S(x)$:
\begin{equation}
\label{messiah}
\left\{
\begin{array}{l}
\displaystyle
\frac{\partial S}{\partial t}+\frac{1}{2m}\biggl(\frac{\partial S}{\partial x}\biggr)^2
+V=\frac{\hbar^2}{2mA}
\frac{\partial^2 A}{\partial x^2}\smallskip\\
\displaystyle m\frac{\partial A}{\partial t}+\frac{\partial A}{\partial x}\frac{\partial S}{\partial x}
+\frac{A}{2}\frac{\partial^2 S}{\partial x^2}=0\\ 
\end{array}
\right.
\end{equation}
From (\ref{messiah}) we see that $S$ and $A$ are {\it coupled} by their equations of motion. 

In classical mechanics if we start from
\begin{equation}
\psi(q,p)=F(q,p) exp[iG(q,p)]
\end{equation}
we can insert it in eq. (\ref{lio1}): $\displaystyle i\frac{\partial\psi}{\partial t}=\HCT\psi$
and, equating the real and imaginary part, we get: 
\begin{equation}
\displaystyle
i\frac{\partial F}{\partial t}=\HCT F, \;\;\;\;\;\;\;\;\;\;\;i\frac{\partial G}{\partial t}=\HCT G
\end{equation}
So we see that in classical mechanics the phase and the modulus {\it decouple} from each other. I owned this analysis
to E. Gozzi \cite{Gozzi3} who has once summarized all this with the sentence:
"{\it What is quantum mechanics? Quantum mechanics is
the  theory of the interaction of a phase with a modulus}". 

As we have just proved, in classical mechanics there
is at least one representation in which this interaction 
is completely lost and the
evolution of the modulus is completely decoupled from the evolution of the phase. 
One may then think that it is useless to deal
with complex wave functions if their phases do not bring in any physical information.  
This is true only if we decide to work with
the Schr\"odinger representation. If, instead, we change representation, as we will do in the next section, 
and use the one where 
$\widehat{p}$ is realized as the derivative
with respect to $\lambda_p$, we shall then show that the mean value of $\widehat{p}$ is related to the phase of the wave functions.
So, if we want to be as general as possible and not just stick to the Schr\"odinger representation, we have to assume
that the classical wave functions are complex objects.

\section{Abstract Hilbert space and $(q,\lambda_p)$ representation}

In the previous section we have restricted ourselves to the Schr\"odinger representation in which both $\widehat{q}$ and 
$\widehat{p}$ are realized as multiplicative operators, and we have worked out everything in this frame. 
What we want to do now is to construct the {\it Hilbert space} of classical mechanics from an {\it abstract} point of view
independently of any particular representation. We can start observing that $\widehat{q}$ and $\widehat{p}$  can be
considered as a complete set of commuting operators whose real eigenvalues can vary with continuity from $-\infty$ to
$+\infty$:
\begin{equation}
\widehat{q}|q,p\rangle=q|q,p\rangle;\;\;\;\;\;\widehat{p}|q,p\rangle=p|q,p\rangle
\end{equation}
The eigenstates $|q,p\rangle$ form an orthonormal and complete set which can be used as a basis for our Hilbert space. The
orthonormality and completeness relations are respectively given by:
\begin{equation}
\displaystyle 
\langle
q^{\prime},p^{\prime}|q^{\prime\prime},p^{\prime\prime}\rangle=
\delta(q^{\prime}-q^{\prime\prime})\delta(p^{\prime}-p^{\prime\prime}),\;\;\;\;\;
\int dq dp\,|q,p\rangle\langle q,p|=1 \label{fortysix}
\end{equation}
The connection between the abstract vectors $|\psi\rangle$ and the wave functions $\psi(q,p)$ is given by the relation 
$\langle q,p|\psi\rangle=\psi(q,p)$. In this basis the operators
$\widehat{q}$ and $\widehat{p}$ are diagonal:
\begin{eqnarray}
&&\langle
q^{\prime},p^{\prime}|\widehat{q}|q^{\prime\prime},p^{\prime\prime}\rangle=q^{\prime}\delta(q^{\prime}-q^{\prime\prime})
\delta(p^{\prime}-p^{\prime\prime});\nonumber\\
&&\langle
q^{\prime},p^{\prime}|\widehat{p}|q^{\prime\prime},p^{\prime\prime}\rangle=p^{\prime}\delta(q^{\prime}-q^{\prime\prime})
\delta(p^{\prime}-p^{\prime\prime})
\end{eqnarray}
while the operators $\displaystyle -i\frac{\partial}{\partial q}\biggl(\displaystyle -i\frac{\partial}{\partial p}\biggr)$
defined by the relations:
\begin{equation}
\displaystyle
\langle
q^{\prime},p^{\prime}\biggl|-i\frac{\partial}{\partial q}\biggl(-i\frac{\partial}{\partial
p}\biggr)\bigg|\psi\rangle=-i\frac{\partial}{\partial q^{\prime}}
\biggl(-i\frac{\partial}{\partial p^{\prime}}\biggr)\langle q^{\prime},p^{\prime}|\psi\rangle \label{fortyeight}
\end{equation}
are self-adjoint. From (\ref{fortysix})-(\ref{fortyeight}) it is easy to check that:
\begin{equation}
\displaystyle
\langle q^{\prime},p^{\prime}\biggl|\biggl[q,-i\frac{\partial}{\partial q}\biggr]\biggr|\psi\rangle=\langle
q^{\prime},p^{\prime}|i|\psi
\rangle,\;\;\;\;\;\;\langle q^{\prime},p^{\prime}\biggl|\biggl[p,-i\frac{\partial}{\partial p}\biggr]\biggr|\psi\rangle=\langle
q^{\prime},p^{\prime}|i|\psi
\rangle \label{fortynine}
\end{equation}
while all other commutators are zero. Because of
the completeness of $\langle q^{\prime},p^{\prime}|$ and the arbitrariness of $|\psi\rangle$ we have that
(\ref{fortynine}) can be turned into the purely operatorial relations:
$\displaystyle \biggl[\widehat{q},-i\frac{\partial}{\partial q}\biggr]=i$ and $\displaystyle
\biggl[\widehat{p},-i\frac{\partial}{\partial p}\biggr]=i$ and so we can identify, from eq. (\ref{comm}),
$\displaystyle
\widehat{\lambda}_q=-i\frac{\partial} {\partial q}$ and 
$\displaystyle \widehat{\lambda}_p=-i\frac{\partial}{\partial p}$.
Now it is easy to show that the Liouville equation (\ref{lio1}) 
is nothing else than a particular representation of the abstract
Liouville equation:
\begin{equation}
\displaystyle
i\frac{\partial}{\partial t}|\psi,t\rangle=\widehat{\lambda}_a\omega^{ab}\partial_bH|\psi,t\rangle \label{ann}
\end{equation}
obtained using as basis the eigenfunctions of $q$ and $p$. 
If we consider as Hamiltonian in the standard phase space the following one: $\displaystyle H=\frac{p^2}{2m}+V(q)$, then 
the Liouville equation (\ref{ann}) becomes:
\begin{equation}
\displaystyle
i\frac{\partial}{\partial t}|\psi,t\rangle=\biggl[\widehat{\lambda}_q\frac{\widehat{p}}{m}-
\widehat{\lambda}_p\partial_qV(q)\biggr]|\psi,t\rangle
\label{anndue}
\end{equation}
Projecting the previous equation onto the basis $\langle q,p|$ we easily obtain:
\begin{eqnarray}
\displaystyle
&&i\frac{\partial}{\partial t}\langle q,p|\psi,t\rangle=\langle q,p\biggl|\widehat{\lambda}_q
\frac{\widehat{p}}{m}\biggr|\psi,t\rangle
-\langle q,p|\widehat{\lambda}_p \partial_qV(q)|\psi,t\rangle=\nonumber\\
&&=-i\frac{\widehat{p}}{m}\frac{\partial}{\partial q}\langle q,p|\psi,t\rangle+i\partial_qV(q)\frac{\partial}{\partial p}\langle
q,p|\psi,t\rangle
\end{eqnarray}
that is equivalent to the usual Liouville equation:
\begin{equation}
\displaystyle
\frac{\partial}{\partial t}\psi(q,p,t)=\biggl[-\frac{p}{m}\frac{\partial}{\partial q}+\partial_qV(q)\frac{\partial}
{\partial p}\biggr]\psi(q,p,t)
\end{equation}

The $|q,p\rangle$ basis is not the only one for the Hilbert space of classical mechanics. A very important
representation
\cite{Abrikosov} is the one in which we consider as basis the simultaneous eigenstates of $\widehat{q}$ and 
$\widehat{\lambda}_p$ which, according to (\ref{comm}), are commuting operators:
\begin{equation}
\widehat{q}|q,\lambda_p\rangle=q|q,\lambda_p\rangle;\;\;\;\;\;\;\;\;\;\;\;
\widehat{\lambda}_p|q,\lambda_p\rangle=\lambda_p|q,\lambda_p\rangle \label{eq2}
\end{equation}
Sandwiching the second relation in (\ref{eq2}) with the bra $\langle q^{\prime},p^{\prime}|$ we obtain
\begin{equation}
\displaystyle
-i\frac{\partial}{\partial p^{\prime}}\langle q^{\prime},p^{\prime}|q,\lambda_p\rangle=\lambda_p\langle q^{\prime},
p^{\prime}|q,\lambda_p\rangle \label{fivefive}
\end{equation}
The solution of this differential equation is:
\begin{equation}
\displaystyle
\langle q^{\prime},p^{\prime}|q,\lambda_p\rangle=\frac{1}{\sqrt{2\pi}}\delta(q-q^{\prime})e^{ip^{\prime}\lambda_p}
\label{fiftysix}
\end{equation}
The states $|q,\lambda_p\rangle$ form a complete set of orthonormal eigenstates, 
i.e. an alternative basis for the
vectors of our classical Hilbert space. In this basis we have:
\begin{equation}
\langle q,\lambda_p|\psi\rangle=\int dq^{\prime} dp\langle q,\lambda_p|q^{\prime},p\rangle\langle q^{\prime},p|\psi\rangle 
\end{equation}
which, via (\ref{fiftysix}), gives:
\begin{equation}
\displaystyle
\psi(q,\lambda_p)=\frac{1}{\sqrt{2\pi}}\int dp \,e^{-ip\lambda_p}\,\psi(q,p) \label{Fou}
\end{equation}
i.e. the wave functions in the new basis are related to the ones in
the Schr\"odinger representation by means of a Fourier transform\footnote{We indicate the wave
functions in the new basis with the same symbol $\psi$ for notational simplicity.}. In this
new representation we have for the $\widehat{p}$ operator:
\begin{eqnarray}
\displaystyle
\langle q,\lambda_p|\widehat{p}|\psi\rangle &=&\int dq^{\prime}dp^{\prime}\langle q,\lambda_p|\widehat{p}|q^{\prime},
p^{\prime}\rangle\langle q^{\prime},p^{\prime}|\psi\rangle=\frac{1}{\sqrt{2\pi}}
\int dp^{\prime}\,p^{\prime}e^{-ip^{\prime}\lambda_p}
\psi(q,p^{\prime})=\nonumber\\
&=&\frac{1}{\sqrt{2\pi}}i\frac{\partial}{\partial\lambda_p}\int dp^{\prime}e^{-ip^{\prime}\lambda_p}\langle
q,p^{\prime}|\psi
\rangle=i\frac{\partial}{\partial\lambda_p}\langle q,\lambda_p|\psi\rangle \label{fiftynine}
\end{eqnarray}
while for $\widehat{\lambda}_p$ we have:
\begin{equation}
\langle q,\lambda_p|\widehat{\lambda}_p|\psi\rangle=\lambda_p\langle q,\lambda_p|\psi\rangle \label{sixty}
\end{equation}
Summarizing (\ref{eq2})-(\ref{sixty}), we can say that in this representation we have to consider 
$\widehat{p}$ as a derivative operator: 
$\displaystyle \widehat{p}=i\frac{\partial}{\partial\lambda_p}$ and $\widehat{\lambda}_p$ 
as a multiplicative one. This is simply a different realization of the usual commutation relation: 
$[\widehat{p},\widehat{\lambda}_p]=i$. 
Using $\langle q,\lambda_p|$ to sandwich eq. (\ref{anndue}) we get that the Louville equation becomes:
\begin{equation}
\displaystyle
i\frac{\partial}{\partial t}\psi(q,\lambda_p)=\frac{1}{m}\frac{\partial}{\partial
q}\frac{\partial}{\partial\lambda_p}\psi(q,\lambda_p) -\lambda_p\partial_qV(q)\psi(q,\lambda_p)
\end{equation}

We shall now show that a lot of the results of the 
previous section were in a certain sense representation-dependent. 
In fact in the new representation, since the momentum $\widehat{p}$ has become an operator of derivation,
we have that the information about its mean value is brought in by the phase of the wave function
similarly to what happens in quantum 
mechanics. 
For example the 
double gaussian state of eq.
(\ref{double}) becomes the following one in the new basis:
\begin{equation}
\displaystyle
\psi(q,\lambda_p,t=0)=\sqrt{\frac{b}{\pi a}}exp\Biggl(-\frac{q^2}{2a^2}\Biggr)
exp\Biggl(-\frac{\lambda_p^2b^2}{2}-ip_i\lambda_p\Biggr) \label{fed1}
\end{equation}
We obtain it by just applying formula (\ref{Fou}) and, 
after the Fourier transform, the wave function which was real in the Schr\"odinger
representation becomes complex. The mean values of $\widehat{q}$ and $\widehat{p}$ are obviously the same
as before:
\begin{equation}
\displaystyle
\bar{q}=\langle\psi|\widehat{q}|\psi\rangle=0,\;\;\;\;\;\;\bar{p}=\langle\psi|\widehat{p}|\psi\rangle=
\langle\psi\biggl|i\frac{\partial}{\partial\lambda_p}\biggr|\psi\rangle=p_i \label{sixtre}
\end{equation}
but now we see that elements appearing in the phases of the wave functions, like $p_i$ in (\ref{fed1}), begin to play an important role since 
they are linked with the mean values of physical observables like $\widehat{p}$. 

Let us now make the evolution of (\ref{fed1}) under the Liouvillian of a free particle.
This Liouvillian in the new representation is given by:
\begin{equation}
\HCT=\frac{1}{m}\frac{\partial^2}{\partial q\partial\lambda_p}
\end{equation}
Its eigenstates associated with the eigenvalues $p\lambda_q/m$ are:
\begin{equation}
\displaystyle
\tau_{\lambda_q,p}(q,\lambda_p)=\frac{1}{2\pi}exp[i\lambda_qq-i\lambda_pp]
\end{equation}
Expanding then the wave function (\ref{fed1}) 
in terms of the $\tau_{\lambda_q,p}$ above and making the evolution of the system, we obtain
at time $t$:
\begin{equation}
\displaystyle
\psi(q,\lambda_p,t)=N\cdot
exp\Biggl[-\frac{q^2}{2a^2}-\frac{p_i^2}{2b^2}-\frac{1}{2}\frac{(\lambda_pma^2b^2
+iqtb^2+ip_ima^2)^2}{a^2b^2(m^2a^2+t^2b^2)}\Biggr] \label{fed2}
\end{equation}
From the previous formula we see how the factor $p_i$ which at time $t=0$ entered only the phase factor,
see (\ref{fed1}), has passed at time $t$ also into the real
part of the wave function, exactly as in the quantum case
we studied before. The expectation values and the variances of $q$ and $p$ are
still given by eqs. (\ref{equno})-(\ref{var12}):
\begin{equation}
\displaystyle
\bar{q}=\int dqd\lambda_p \;q\cdot|\psi(t)|^2=\frac{p_it}{m},\;\;\;\;\;\;\bar{p}=p_i
\end{equation}
and:
\begin{equation}
\displaystyle
\overline{(\Delta q)^2}=\frac{a^2}{2}+\frac{b^2}{2}\frac{t^2}{m^2},\;\;\;\;\;\;\overline{(\Delta p)^2}=\frac{b^2}{2}
\end{equation}
This is so since they are observable quantities and, consequently, they are independent of the representation we are using.
In the new representation making a wave function of the form (\ref{fed1})
well-localized both in $q$ and in $\lambda_p$ means sending $a\to 0,b\to\infty$. In this
limiting case  we have that $\overline{(\Delta q)^2}\to\infty$ at every instant of time $t>0$ and so there is a sort of
spreading of the wave function. This is not surprising. In fact if the initial wave function is very peaked around
$q=\lambda_p=0$ then we know precisely the values of $q$ and $\lambda_p$, instead of the values of $q$ and $p$. From the
commutator $[\widehat{p},\widehat{\lambda}_p]=i$  we can derive the following uncertainty relation $\displaystyle
\Delta p\cdot\Delta\lambda_p\ge 1/2$, where $\Delta p$ and $\Delta\lambda_p$ are the square roots of the 
mean square deviations.
So, if we determine with absolute precision
$\lambda_p$, as we do with the limit $b\to\infty$, the momentum
$p$ is completely undetermined. Consequently also the position $q$ at every instant $t>0$ 
is completely undetermined, because it follows the classical equations of motion 
$\dot{q}=p/m$. This fact has, as
an immediate consequence, the spreading of $q$ and the complete delocalization of the wave function at every instant following the
initial one.\footnote{The usual mechanics of the single particle can be reproduced also in this representation but we have to take
the limit
$a\to 0,b\to 0$, i.e. we have to use for $\lambda_p$ a plane wave of the type $exp(-ip_i\lambda_p)$.}

Another aspect that we can study is the continuity equation. We have seen in the second
section that the continuity equation in the Schr\"odinger representation is nothing else than the usual Liouville equation for
the probability density $\rho$. What happens in the other representation we have studied in eqs.
(\ref{eq2})-(\ref{sixty})? According to what we have already
seen, the Liouville equation for $\psi$ is:
\begin{equation}
\displaystyle
i\frac{\partial}{\partial t}\psi(q,\lambda_p)=\frac{1}{m}\frac{\partial}{\partial
q}\frac{\partial}{\partial\lambda_p}\psi(q,\lambda_p) -\lambda_pV^{\prime}(q)\psi(q,\lambda_p) \label{seinove}
\end{equation}
while the one for the $\psi^*(q,\lambda_p)$ is the complex conjugate:
\begin{equation}
\displaystyle
-i\frac{\partial}{\partial t}\psi^*(q,\lambda_p)=\frac{1}{m}\frac{\partial}{\partial q}\frac{\partial}{\partial\lambda_p}
\psi^*(q,\lambda_p)-\lambda_pV^{\prime}(q)\psi^*(q,\lambda_p) \label{settezero}
\end{equation}
From (\ref{seinove}) and (\ref{settezero}) we can obtain the equation for $\rho(q,\lambda_p)=\psi^*(q,\lambda_p)
\psi(q,\lambda_p)$. It is of the form:
\begin{equation}
\displaystyle
\frac{\partial}{\partial t}\rho(q,\lambda_p)+J=0
\end{equation}
where:
\begin{equation}
\displaystyle
J=\frac{i}{m}\bigg(\psi^*\frac{\partial}{\partial q}\frac{\partial}{\partial\lambda_p}\psi-
\psi\frac{\partial}{\partial q}\frac{\partial}{\partial\lambda_p}\psi^*\bigg)
\end{equation}
So we notice that in this case $\rho(q,\lambda_p)$ evolves with an equation that is completely 
different from the Liouville equation and there is no manner to write $J$ in terms only of $\rho$.
Moreover, if we write 
$\psi(q,\lambda_p)=\sqrt{\rho}\,exp[iS(q,\lambda_p)]$,
the phase $S(q,\lambda_p)$ will enter explicitly $J$ and the equation of $\rho$, i.e. we
have a situation very similar to that of quantum mechanics
where phases and modulus are coupled in the equations of motion. 

Another aspect that the $(q,\lambda_p)$ representation of classical mechanics and the standard quantum
one have in common is that, even if we prepare a real wave function of $q$ and $\lambda_p$ at the initial time $t=0$, phases
will be created in general by $\HCT$ during the evolution. This can be seen by means of our usual example. In fact, if we put
$p_i=0$, we have from eqs. (\ref{fed1}) and (\ref{fed2}) that:
\begin{equation}
\displaystyle\sqrt{\frac{b}{\pi a}}exp\biggl(-\frac{q^2}{2a^2}-\frac{\lambda_p^2b^2}{2}\biggr)
\longrightarrow N\cdot
exp\biggl[-\frac{q^2}{2a^2}-\frac{1}{2}\frac{(\lambda_pma^2b^2+iqtb^2)^2}{a^2b^2(m^2a^2+t^2b^2)}
\biggr]
\end{equation}
that is:
\begin{equation}
\HCT: \;\;\;\psi(q,\lambda_p,t=0\;\; without\; phases)\;\;\;\longrightarrow\;\;\;\psi(q,\lambda_p,t \;\;with \;phases)
\end{equation}
Since the $(q,\lambda_p)$ representation has all these features in common with quantum mechanics it is not a case
that this representation turns out to be \cite{Abrikosov} the one where the transition to quantum mechanics is best 
understood.

\section{Two-slit experiment}

Having formulated classical
mechanics in the same mathematical language of quantum mechanics, we think it may be useful to analyze the two-slit
experiment in the classical KvN formalism and compare it with its quantum analogue. 
This kind of experiment is central in quantum mechanics and its mystery is best summarized in these words of
Feynman \cite{Feynman}: {\it "The question is, how does [the two-slit experiment] really work? What
machinery is actually producing this thing? Nobody knows any machinery. The mathematics can be made more precise;
you can mention that they are {\it complex} numbers, and a couple of other minor points which have nothing to 
do with the main idea. But the deep mystery is what I have described, and no one can go any deeper today"}.
As Feynman mention in the lines above one could think that the interference effects are there because of the {\it complex}
nature of the wave functions. Then it is natural to check what happens in the classical KvN
case where, as we showed in the previous section, the wave functions have to be complex. We will actually show
that, despite the complex nature of these wave functions, interference effects do not appear. This confirms,
as Feynman suspected, that the mystery of quantum mechanics is deeper than that. 

If we want to describe a classical two-slit experiment we have to
deal with a two dimensional problem. Let us call
$y$ the axis along which our beam propagates and
$x$ the orthogonal axis. We suppose that $y=0$ is the starting coordinate of our beam. 
The centers of the two slits $\Delta_1$ and $\Delta_2$ are placed respectively at $x_{\scriptscriptstyle A}$ and
$-x_{\scriptscriptstyle_A}$ on a first plate which has coordinate $y_{\scriptscriptstyle F}$ along the $y$ axis. The
final screen is placed at
$y_{\scriptscriptstyle S}$ like in the figure below. 
\bigskip

\begin{center}
\begin{picture}(248,80)
\put(0,20){\makebox(0,0){$\scriptstyle{-x_A}$}}
\put(0,60){\makebox(0,0){$\scriptstyle{x_A}$}}
\put(16,80){\line(0,-1){80}}
\put(16,20){\makebox(0,0){$\bullet$}}
\put(16,60){\makebox(0,0){$\bullet$}}
\put(24,0){\makebox(0,0){$\scriptstyle{0}$}}
\put(230,80){\line(0,-1){80}}
\put(123,80){\line(0,-1){18}}
\put(120,62){\line(1,0){6}}
\put(120,58){\line(1,0){6}}
\put(123,58){\line(0,-1){36}}
\put(120,22){\line(1,0){6}}
\put(120,18){\line(1,0){6}}
\put(123,18){\line(0,-1){18}}
\put(135,20){\makebox(0,0){$\scriptstyle{\Delta_2}$}}
\put(135,60){\makebox(0,0){$\scriptstyle{\Delta_1}$}}
\put(135,0){\makebox(0,0){$\scriptstyle{y_F}$}}
\put(242,0){\makebox(0,0){$\scriptstyle{y_S}$}}
\end{picture}
\end{center}

\bigskip
To simplify the problem we make the assumption\footnote{A similar assumption was made by Feynman in his analysis
of the diffraction experiment \cite{Hibbs}.} that the motion of the particle along the $y$
direction is known precisely. This means that at the initial time we know with absolute precision the position and the
momentum of the particle, for example $y(0)=0,\;p_y(0)=p_y^0$. With this prescription we are sure that the beam will 
arrive at
the two slits after a time $t_{\scriptscriptstyle F}=y_{\scriptscriptstyle F}m/p_y^0$ and at the
final screen after a time 
$t_{\scriptscriptstyle S}=y_{\scriptscriptstyle S}m/p_y^0$. In this way we can concentrate
ourselves only on the behaviour of the particles along the $x$-axis.
Suppose we consider, along $x$, a double-gaussian wave function
with an arbitrary phase factor $G(x,p_x)$:
\begin{equation}
\displaystyle
\psi(x,p_x,t=0)=\frac{1}{\sqrt{\pi ab}}exp\biggl[-\frac{x^2}{2a^2}-\frac{p_x^2}{2b^2}+iG(x,p_x)\biggr]\label{dunia}
\end{equation}
We assume $a$ and $b$ sufficiently large, i.e. the initial classical wave function sufficiently spread, 
in order to allow the beam to arrive at both slits. The evolution of the wave
function will be via the free kernel of propagation (\ref{kernel}) up to the time 
$t_{\scriptscriptstyle F} =y_{\scriptscriptstyle F}m/p_y^0$ that is when the beam arrives at the first plate. 
The wave function at the time $t_{\scriptscriptstyle F}$ will be:
\begin{equation}
\displaystyle
\psi(x,p_x,t_{\scriptscriptstyle F})=\frac{1}{\sqrt{\pi
ab}}exp\biggl[-\frac{1}{2a^2}\biggl(x-\frac{p_xy_{\scriptscriptstyle F}}{p_y^0}\biggr)^2-\frac{p_x^2}{2b^2}
+iG\biggl(x-\frac{p_xy_{\scriptscriptstyle F}}{p_y^0},p_x\biggr)\biggl] \label{dunia2}
\end{equation}
Let us suppose that the width of the two slits is $2\delta$, then 
the particles which at time $t_{\scriptscriptstyle F}$ are outside of the
two intervals $\Delta_1=(x_{\scriptscriptstyle A}-\delta, x_{\scriptscriptstyle A}+\delta)$ and
$\Delta_2=(-x_{\scriptscriptstyle A}-\delta, -x_{\scriptscriptstyle A}+\delta)$ are absorbed by the first plate and they 
don't
arrive at the final screen at all. Using Feynman's words again: "{\it All particles which miss the slit[s] are captured and
removed from the experiment} {\cite{Hibbs}". Therefore the wave function just after
$t_{\scriptscriptstyle F}$ can be rewritten in a compact way using a series of $\theta$-Heavyside functions:
\begin{equation}
\displaystyle 
\psi(x,p_x,t_{\scriptscriptstyle F}+\epsilon)=N\,
\psi(x,p_x,t_{\scriptscriptstyle F})\,[C_1(x)+C_2(x)] \label{mar}
\end{equation}
where 
$C_1(x)=\theta(x-x_{\scriptscriptstyle A}+\delta)-\theta(x-x_{\scriptscriptstyle
A}-\delta)$ is the function that parametrizes the slit $\Delta_1$,
$C_2(x)=\theta(x+x_{\scriptscriptstyle A}+\delta)-\theta(x+x_{\scriptscriptstyle
A}-\delta)$ is the one that parametrizes the slit $\Delta_2$ and $N$ is a suitable normalization factor
chosen in such a way that:
$\int dxdp_x \,|\psi(x,p_x,t_{\scriptscriptstyle F}+\epsilon)|^2=1$. 
Beyond the double slit we will propagate 
the $\psi$ of eq. (\ref{mar}). With our choice of the cut-off functions $C_1$ and $C_2$, 
at $t_{\scriptscriptstyle F}+\epsilon$ the wave function 
$\psi$ is different from $0$ only  if $x\in\Delta_1\;or\; x\in\Delta_2$.
Since there is no limitation in
the momentum along the $x$-axis we expect that the wave function $\psi$ will spread 
along $x$ and, while time passes, it will become different from zero
also outside the intervals $\Delta_1$ and $\Delta_2$. 

Using the kernel of evolution for free particles (\ref{kernel}) 
we can obtain from (\ref{mar}) the wave function at time $t_{\scriptscriptstyle S}
=y_{\scriptscriptstyle S}m/p_y^0$ that is the time when the beam arrives at the final screen is:
\begin{eqnarray}
\displaystyle
\psi(x,p_x,t_{\scriptscriptstyle S})&=&N \cdot
exp\biggl[-\frac{1}{2a^2}\biggl(x-\frac{p_xy_{\scriptscriptstyle
S}}{p_y^0}\biggr)^2-\frac{p_x^2}{2b^2}\biggl]\cdot exp\biggl[iG\biggl(x-\frac{p_xy_{\scriptscriptstyle S}}
{p_y^0},p_x\biggr)\biggr]\nonumber\\ &&\cdot
\{C_1(x-\bar{a}p_x)
+C_2(x-\bar{a}p_x)\} \label{marfin}
\end{eqnarray}
where $ \bar{a}=(y_{\scriptscriptstyle S}-y_{\scriptscriptstyle F})/p_y^0$.
The probability density to find a particle in a certain point $x$ on the last screen 
\begin{equation}
P(x)=\int_{-\infty}^{\infty}dp_x|\psi(x,p_x,t_{\scriptscriptstyle S})|^2 \label{pi}
\end{equation}
We have to integrate over $p_x$ because we are interested in the number of
particles that arrive at the final plate independently of their momentum.
At this point we notice a first important property: even starting from an initial wave function
with an arbitrary phase factor $G(x,p_x)$, at time $t_{\scriptscriptstyle S}$ we have for the entire wave function
a common phase factor of the form 
$\displaystyle G\biggl(x-\frac{p_xy_{\scriptscriptstyle S}}{p_y^0},p_x\biggr)$, see eq. (\ref{marfin}). 
So $G$ will disappear completely in the evaluation of the modulus square and, consequently,
in the $P(x)$ of eq. (\ref{pi}). Therefore the phase $G$ of the initial wave function (\ref{dunia})
cannot have any observable consequence in the figure on the final screen.

The second important thing to notice is that, because of the properties of the $\theta$-functions,
we have that the cut-off term $C_1+C_2$ in (\ref{marfin}) is idempotent:
\begin{equation}
\displaystyle
(C_1+C_2)^2=C_1+C_2
\end{equation}
Therefore we can rewrite (\ref{pi}) as:
\begin{eqnarray}
P(x)&=&\int_{-\infty}^{\infty}dp_x|\psi(x,p_x,t_{\scriptscriptstyle S})|^2
=N\cdot\biggl[\int_{-\infty}^{\infty}dp_xF^2(x,p_x,t_{\scriptscriptstyle S})C_1(x-\bar{a}p_x)+\nonumber\\
&&+\int_{-\infty}^{\infty}dp_xF^2(x,p_x,t_{\scriptscriptstyle S})C_2(x-\bar{a}p_x)\biggr] \label{rui}
\end{eqnarray}
where $N$ is a normalization factor and $F$ is given by:
\begin{equation}
\displaystyle
F(x,p_x, t_{\scriptscriptstyle S})\equiv exp\biggl[-\frac{1}{2a^2}\biggl(x-\frac{p_xy_{\scriptscriptstyle
S}}{p_y^0}\biggr)^2-\frac{p_x^2}{2b^2}\biggr] \label{intcla}
\end{equation}

Let us now re-arrange the arguments inside the $\theta$-functions of the $C_1$ and the $C_2$ as follows:
\begin{eqnarray}
\displaystyle
&&C_1(x-\bar{a}p_x)=\theta\biggl(-p_x+\frac{x-x_{\scriptscriptstyle
A}+\delta}{\bar{a}}\biggr)-\theta\biggl(-p_x+\frac{x-x_{\scriptscriptstyle
A}-\delta}{\bar{a}}\biggr)\nonumber\\
&&C_2(x-\bar{a}p_x)=\theta\biggl(-p_x+\frac{x+x_{\scriptscriptstyle
A}+\delta}{\bar{a}}\biggr)-\theta\biggl(-p_x+\frac{x+x_{\scriptscriptstyle
A}-\delta}{\bar{a}}\biggr)
\end{eqnarray} 
Remembering the properties of the $\theta$-Heavyside functions it is easy to realize that when $p_x$ is not in one
of the two intervals:
$\displaystyle D_1=\biggl[\frac{x-x_{\scriptscriptstyle A}-\delta}{\bar{a}}, \frac{x-x_{\scriptscriptstyle
A}+\delta}{\bar{a}}\biggr]$ or
$\displaystyle D_2=\biggl[\frac{x+x_{\scriptscriptstyle A}-\delta}{\bar{a}}, \frac{x+x_{\scriptscriptstyle
A}+\delta}{\bar{a}}\biggr]$ there is no contribution to the modulus square. 
Therefore, apart from the normalization coefficient $N$, we have 
that the final plot $P(x)$ given by eq. (\ref{rui}) can be written as:
\begin{equation}
\displaystyle
P(x)=\int_{-\infty}^{\infty}dp_x\, |\psi(x,p_x,t_{\scriptscriptstyle S})|^2=
N\cdot\biggl[\int_{D_1} dp_x\,F^2(x,p_x,t_{\scriptscriptstyle S})+\int_{D_2}
dp_x\,F^2(x,p_x,t_{\scriptscriptstyle S})\biggr]
\label{imp}
\end{equation}
where $F$ is the function of eq. (\ref{intcla}).

Now let us keep open only the first slit $\Delta_1$ and repeat the previous calculations. We can propagate
the initial wave function (\ref{dunia}) up to the time $t_{\scriptscriptstyle F}$ when the system is again described
by the $\psi(x,p_x,t_{\scriptscriptstyle F})$ of eq. (\ref{dunia2}). The difference is that now we have to parametrize only
the first slit $\Delta_1$. Therefore the second cut-off function $C_2$ is identically zero. Since $C_1$ itself 
is an idempotent function we can repeat the same steps as before, eqs. (\ref{mar})-(\ref{imp}),
freezing everywhere $C_2$ to 
zero. As final result we obtain the following probability on the last screen:
\begin{equation}
P(x)=K\int_{D_1}dp_xF^2(x,p_x,t_{\scriptscriptstyle S}) \label{imp1}
\end{equation}
where $F$ is again given by eq. (\ref{intcla}). In the same manner keeping open only the slit $\Delta_2$
we will obtain that:
\begin{equation}
P(x)=K\int_{D_2}dp_xF^2(x,p_x,t_{\scriptscriptstyle S}) \label{imp2}
\end{equation}
So, comparing (\ref{imp}) with (\ref{imp1}) and (\ref{imp2}), it is clear that when we keep
open both slits $\Delta_1+\Delta_2$ {\it the total probability is the sum of the probabilities} of
having kept open first the slit $\Delta_1$ and then the slit $\Delta_2$. The first integral in (\ref{imp})
is then the probability for the particle to pass through the slit $\Delta_1$, while the second integral
is the probability to pass through the slit $\Delta_2$. 
So, even if we start from complex wave functions in 
the classical Hilbert space,
every interference effect disappears. This is very clear from
Figure \ref{classical} which shows the plot of the $P(x)$ of eq. (\ref{imp}) with the particular numerical values
$\displaystyle y_{\scriptscriptstyle S}/p_y^0=2,\;a=b=1, x_{\scriptscriptstyle A}=1,\;\delta=0.1$.

We will now perform a similar calculation at the quantum level and compare it with the previous 
classical experiment. In order to get an analytic result we will do a simplification, that is 
we will assume that the motion along $y$ is the same classical motion analyzed before. The reason for this
assumption is that otherwise we would not be able to determine the time $t_{\scriptscriptstyle F}$
at which the wave function arrives on the plate with the two slits.
Along $x$ instead we will assume that the motion is fully quantum mechanical. So our overall approach to the 
quantum case is actually a "semiclassical" approach. Nevertheless this will be sufficient to see 
the difference with the purely classical case we have analyzed previously. 

In our semiclassical approach at the initial time the system 
along the $y$-axis is described by a double Dirac delta
$\delta(y)\delta(p_y-p_y^0)$ and this double Dirac delta evolves in time with the Liouvillian. In this way we know 
which is the time the particles arrive at the two slits. It is exactly the same as before. 
Along the other axis, $x$, we consider instead an initial wave function given by:
\begin{equation}
\displaystyle
\psi(x)=\sqrt{\frac{1}{\sqrt{\pi}a}}exp\biggl(-\frac{x^2}{2a^2}\biggr) \label{binn}
\end{equation}
With this choice the mean value of both $x$ and $p_x$ is zero at the initial time as in the classical case described by eq.
(\ref{dunia}). Making the above wave function evolve in time via the quantum Schr\"odinger operator, at time
$t_{\scriptscriptstyle F}$ we obtain:
\begin{equation}
\displaystyle
\psi(x,t_{\scriptscriptstyle F})=\sqrt{\frac{ma}{\sqrt{\pi}(ma^2+i\hbar t_{\scriptscriptstyle F})}}
exp\biggl[-\frac{1}{2}\frac{mx^2}{ma^2+i\hbar t_{\scriptscriptstyle F}}\biggr]
\end{equation}

Let us parametrize the two slits by means of the same series of $\theta$-Heavyside functions 
we have used in the classical case:
\begin{equation}
C_1(x)=\theta(x-x_{\scriptscriptstyle A}+\delta)-\theta(x-x_{\scriptscriptstyle A}-\delta),\;\;\;\;
C_2(x)=\theta(x+x_{\scriptscriptstyle A}+\delta)-\theta(x+x_{\scriptscriptstyle A}-\delta)
\end{equation}
Just after the wave function has passed the plate with the two slits we have that:
\begin{equation}
\displaystyle
\psi(x,t_{\scriptscriptstyle F}+\epsilon)=\bar{N}\cdot exp\biggl(-\frac{1}{2}\frac{mx^2}{ma^2+i\hbar t_{\scriptscriptstyle
F}}\biggr)\bigl[C_1(x)+C_2(x)\bigr]
\end{equation}
Using now the kernel of propagation \cite{Hibbs} for a quantum free particle which is given by:
\begin{equation}
\displaystyle
K(x_b,t_b|x_a,t_a)=\biggl[\frac{2\pi i\hbar(t_b-t_a)}{m}\biggr]^{-1/2}exp\;\frac{im(x_b-x_a)^2}{2\hbar(t_b-t_a)}
\end{equation}
we get that at time $t_{\scriptscriptstyle S}$ the wave function is:
\begin{eqnarray}
\displaystyle
&&\psi(x,t_{\scriptscriptstyle S})=\bar{N}_1\int_{-\infty}^{+\infty}dx_{\scriptscriptstyle F}
\,exp\biggl[\frac{im(x-x_{\scriptscriptstyle
F})^2}{2\hbar(t_ {\scriptscriptstyle S}-t_{\scriptscriptstyle F})}-\frac{mx_{\scriptscriptstyle F}^2}{2(ma^2+
i\hbar t_{\scriptscriptstyle F})}\biggr][C_1(x_{\scriptscriptstyle F})+C_2(x_{\scriptscriptstyle F})]\nonumber\\
\label{psifin}
\end{eqnarray}
where $\bar{N}_1$ is a new normalization constant.
Differently from the classical case, the quantum kernel of propagation is not a simple Dirac delta and
the previous integral cannot be done explicitly. Anyway we can employ the properties
of the $\theta$-functions in order to rewrite (\ref{psifin}) as:
\begin{eqnarray}
\displaystyle
\psi(x,t_{\scriptscriptstyle S})&=&\bar{N}_1\biggl\{\int_{x_{\scriptscriptstyle A}-\delta}^{x_{\scriptscriptstyle A}+\delta}
dx_{\scriptscriptstyle F}\,exp \biggl[\frac{im(x-x_{\scriptscriptstyle F})^2}
{2\hbar(t_{\scriptscriptstyle S}-t_{\scriptscriptstyle
F})} -\frac{mx_{\scriptscriptstyle F}^2}{2(ma^2+
i\hbar t_{\scriptscriptstyle F})}\biggr]+\nonumber\\
&&+\int_{-x_{\scriptscriptstyle A}-\delta}^{-x_{\scriptscriptstyle A}+\delta}
dx_{\scriptscriptstyle F}\,exp\biggl[\frac{im(x-x_{\scriptscriptstyle F})^2}
{2\hbar(t_{\scriptscriptstyle S}-t_{\scriptscriptstyle
F})} -\frac{mx_{\scriptscriptstyle F}^2}{2(ma^2+
i\hbar t_{\scriptscriptstyle F})}\biggr]\biggr\} \label{previous}
\end{eqnarray}
In (\ref{previous}) we have two integrals of the {\it same} function over the two {\it different} intervals
$\Delta_1=(x_{\scriptscriptstyle A}-\delta,x_{\scriptscriptstyle A}+\delta)$ and 
$\Delta_2=(-x_{\scriptscriptstyle A}-\delta,-x_{\scriptscriptstyle A}+\delta)$. 
The results will be two complex numbers $\psi_1$ and $\psi_2$ with 
{\it different phases}. So, differently from the classical case (\ref{marfin}), the quantum wave function
on the final screen $\psi(x,t_{\scriptscriptstyle S})$ has not a common phase factor and as a consequence the 
relative phases of $\psi_1$ and $\psi_2$ will play a crucial role in giving interference effects. In fact if we
re-write the final wave function as:  
\begin{equation}
\psi(x, t_{\scriptscriptstyle S})=\bar{N}_1\,
[\psi_1(x, t_{\scriptscriptstyle S})+\psi_2(x, t_{\scriptscriptstyle S})]
\end{equation}
the probability on the last screen 
is given by the modulus square of $\psi(x, t_{\scriptscriptstyle S})$:
\begin{equation}
P(x, t_{\scriptscriptstyle S})=|\psi_1(x, t_{\scriptscriptstyle S})|^2+
|\psi_2(x, t_{\scriptscriptstyle S})|^2+\psi_1^*(x, t_{\scriptscriptstyle S})\psi_2(x, t_{\scriptscriptstyle S})
+\psi_1(x, t_{\scriptscriptstyle S})\psi_2^*(x, t_{\scriptscriptstyle S})
\end{equation}
Note that the last two terms in the previous formula are not identically zero.
If we make a plot of $P(x, t_{\scriptscriptstyle S})$ as a function of $x$ we 
see the evidence of the interference typical of quantum mechanics with a central
maximum and a series of secondary maxima. This can be seen from 
Figure \ref{quantum2} which is the plot of $P(x)$ in the case $t_{\scriptscriptstyle S}=2,
t_{\scriptscriptstyle F}=1, \break
m=a=1,\hbar=1,\delta=0.1$ for two different distances of the slits: $2x_{\scriptscriptstyle A}=1$ and $2x_{\scriptscriptstyle A}=2$
respectively. 
We can note the presence of six minima in the first case and of twelve minima in the second one. This in perfect
agreement with the well-known relation that the distance $\Delta x$ between two successive maxima or minima in an
interference figure is inversely proportional to the distance $2x_{\scriptscriptstyle A}$ between the slits.
Therefore, even considering a quantum evolution only along the
$x$-axis, the quantum wave functions create interference effects and the final result is a series of
maxima and minima, like in the real experiment. 

Summarizing the results of this section we can say that if we make the evolution along the $x$ axis with 
the Schr\"odinger Hamiltonian $\widehat{H}$, even starting from a real wave function, 
like (\ref{binn}), phases will appear
during the evolution in a non trivial way and they will contribute to create interference effects.
Instead in the evolution along $x$ with the Liouvillian $\HCT$, even starting from a complex wave function, 
like (\ref{dunia}),
the phase appears as a common factor for the entire $\psi$ on the final screen and so it does not 
contribute to $|\psi|^2$ and it has not observable consequences. 
We feel that the two different behaviours are basically due to 
the different forms of the evolution operators in the classical
and in the quantum case.

\section{Conclusions}

In this paper we have shown, by means of simple and pedagogical examples, some of the differences between the
operatorial approaches to classical and quantum mechanics. While in quantum mechanics the phases and the modulus
of the wave functions are always coupled, eq. (\ref{messiah}), in classical mechanics we can find a representation,
the $(q,p)$ one, where we can decouple them. Their being coupled in some representation of classical 
mechanics, like 
the $(q,\lambda_p)$ one, is only an {\it apparent} phenomenon. We think that this is the real profound 
feature of quantum mechanics which makes it different from classical mechanics: that in quantum mechanics there is
no way to decouple the phases from the modulus by just going to a proper representation like we can do in
classical mechanics \cite{Gozzi3}. Of course we have not given a general proof of this in the sense 
that we have not really proven that in quantum mechanics this proper representation does not exist, but we strongly
feel this is the case \cite{last}. 

In this paper we have also performed the classical analog of the two-slit experiment and we have seen that interference
effects do not appear. We have proved that by performing the detailed calculations giving the motion
of the classical "wave functions" in the two-slit experiment. That was just a particular example and we do not know 
if it works in the same way in general like, for example, in the multiple-slit case or in other phenomena where at 
the quantum level there is interference. To do the detailed calculations in all these cases may turn out to be
even more difficult than in the two-slit case. To by-pass those calculations we would like to find 
a general proof of the absence of interference effects in the KvN formalism. For sure in this general 
proof a crucial role will be played by the different form that the classical $\HCT$ and the quantum $\widehat{H}$ have.
Another crucial role may be played by the {\it universal} symmetries present in classical mechanics
and discovered in \cite{Gozzi2}. Those symmetries may be trigger a sort of superselection
mechanism which may prevent the interference from appearing. All these features may be also those that,
being absent in quantum mechanics, forbid the decoupling of the phase and the modulus. Work is
in progress \cite{last} on these problems.

\appendix
\makeatletter
\@addtoreset{equation}{section}
\makeatother
\renewcommand{\theequation}{\thesection.\arabic{equation}}

\section{Appendix }

In this appendix we will show that there is no contradiction in the postulate of KvN
of having for $\psi$ the same evolution as for $\rho=|\psi|^2$.
The kernel $P(\varphi, t|\varphi_i,t_i)$ for $\rho$ is just (\ref{prob}) which is a Dirac delta.
So, postulating this to be the same as the kernel $K(\varphi,t|\varphi_i,t_i)$ of propagation of
the $\psi$, in the case of a free particle we have:
\begin{equation}
\displaystyle
K(\varphi,t|\varphi_i,t_i=0)=\delta\biggl(q-q_i-\frac{p_it}{m}\biggr)\delta(p-p_i)
\end{equation}
Let us now use this expression to check what we get for the kernel of $\rho$ knowing that $\rho(t)=|\psi(t)|^2$.
\begin{eqnarray}
\displaystyle
\rho(t)=\psi^*(t)\psi(t)&=&\int d\varphi_iK^*(\varphi,t|\varphi_i,0)\psi^*(\varphi_i,0)\cdot\int
d\varphi_i^{\prime}K(\varphi,t|\varphi_i^{\prime},0)\psi(\varphi_i^{\prime},0)=
\nonumber\\ 
&=&\int
dq_idp_i\delta\biggl(q-q_i-\frac{p_it}{m}\biggr)\delta(p-p_i)\psi^*(q_i,p_i,0)\cdot
\nonumber\\
&&\cdot\int
dq^{\prime}_idp^{\prime}_i\delta\biggl(q-q^{\prime}_i-\frac{p^{\prime}_it}{m}\biggr)\delta(p-p^{\prime}_i)
\psi(q^{\prime}_i,p^{\prime}_i,0) \label{diruno}
\end{eqnarray}
Now we can use the properties of the Dirac deltas to rewrite:
\begin{eqnarray}
\rho(t)&=&\int dq_idp_idq_i^{\prime}dp_i^{\prime}\delta\biggl(q-q_i-\frac{p_it}{m}\biggr)\delta(p-p_i)
\delta(p_i-p_i^{\prime})\cdot\nonumber\\
&&\;\;\;\cdot\delta\biggl(q_i-q_i^{\prime}+(p_i-p_i^{\prime})\frac{t}{m}\biggr)\psi^*(q_i,p_i,0)\psi(q_i^{\prime},
p_i^{\prime},0) \label{dirdue}
\end{eqnarray}
The integrals over the primed variables can be done explicitly:
\begin{equation}
\int dq_i^{\prime}dp_i^{\prime}\,\delta(p_i-p_i^{\prime})
\delta\biggl(q_i-q_i^{\prime}+(p_i-p_i^{\prime})\frac{t}{m}\biggr)\psi^*(q_i,p_i,0)\psi(q_i^{\prime},
p_i^{\prime},0)=\rho(\varphi_i,0) \label{dirtre}
\end{equation}
Substituting (\ref{dirtre}) into (\ref{dirdue}) we have finally:
\begin{equation}
\rho(t)=\int dq_idp_i K(\varphi, t|\varphi_i,0)\rho(\varphi_i,0)
\end{equation}
From this relation we get that the kernel of propagation of the $\rho$ is the same as the one
of the $\psi$ and this proves that there is no contradiction in the KvN postulate.

\section*{Acknowledgments}

I would like to thank M. Reuter, E. Deotto and L. Marinatto for many helpful discussions. A very special thank to  
E. Gozzi for a lot of helpful suggestions and for having inspired with his ideas the greatest part of this work. 
This research has been supported by grants from INFN, MURST and the University of Trieste.

\newpage

\noindent {\LARGE \underline{Figure Caption}}

\bigskip
\bigskip
\bigskip

\noindent {\large Fig. 1:} Classical Two-Slit Experiment.\\

\noindent {\large Fig. 2:} Quantum Two-Slit Experiment.

\newpage

\begin{center}
\begin{figure}
\includegraphics{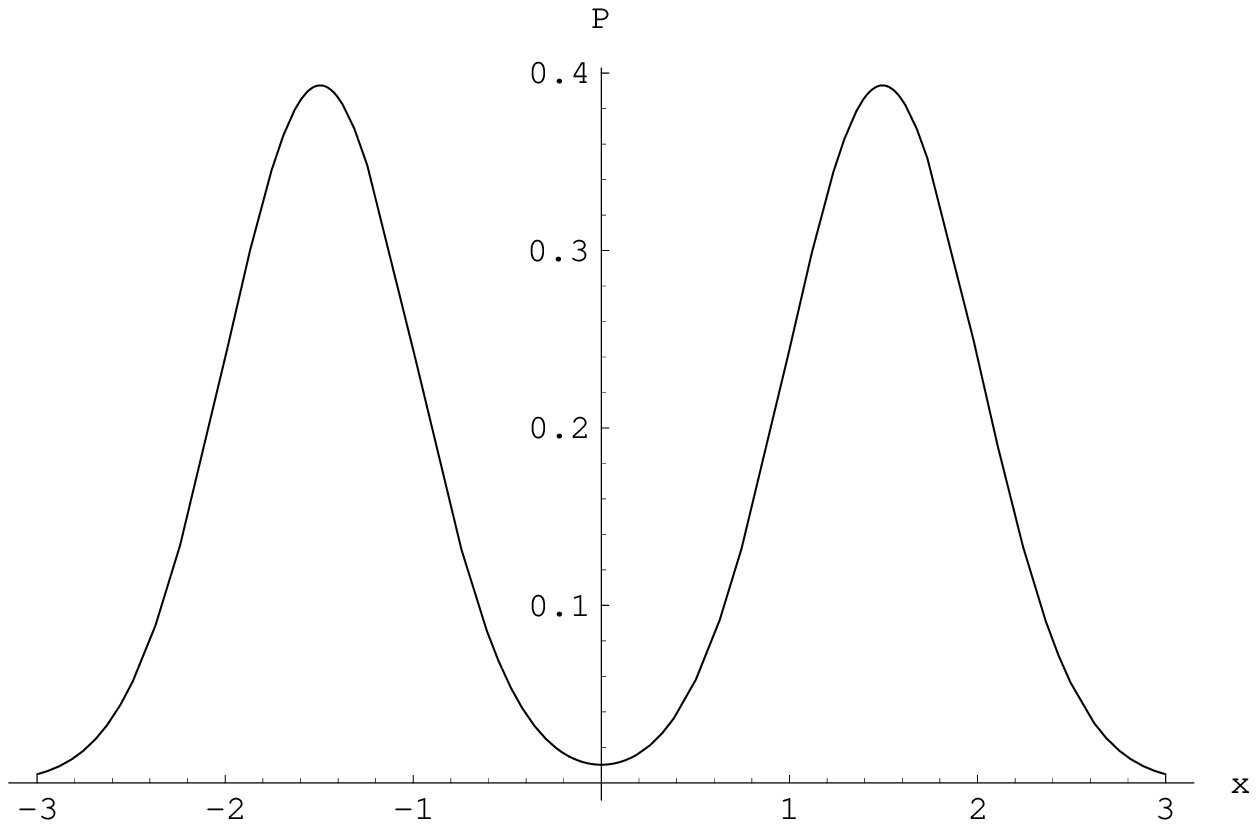}
\caption{Classical Two-Slit Experiment}\label{classical}
\end{figure}
\end{center}

\begin{center}
\begin{figure}
\includegraphics{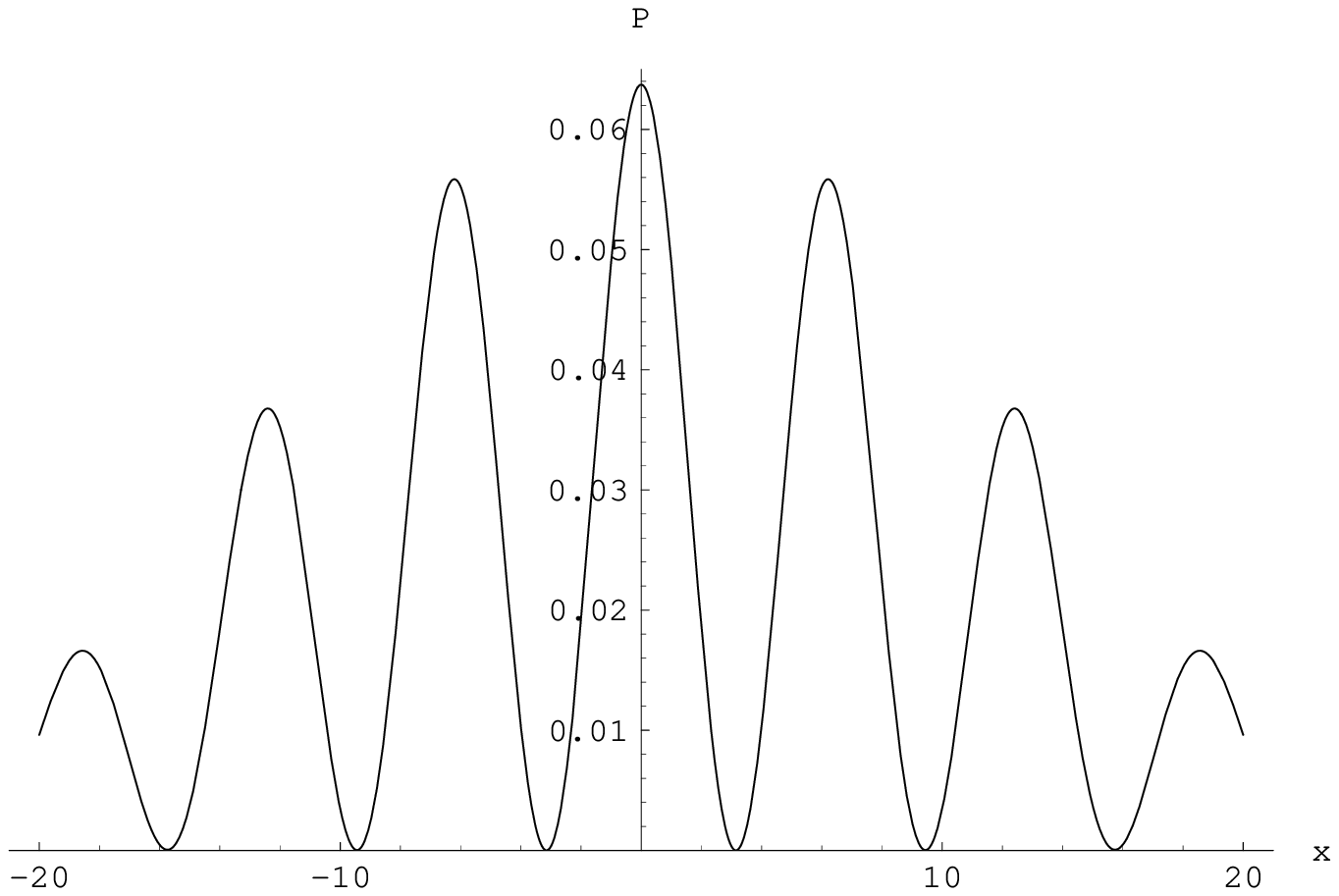}

\includegraphics{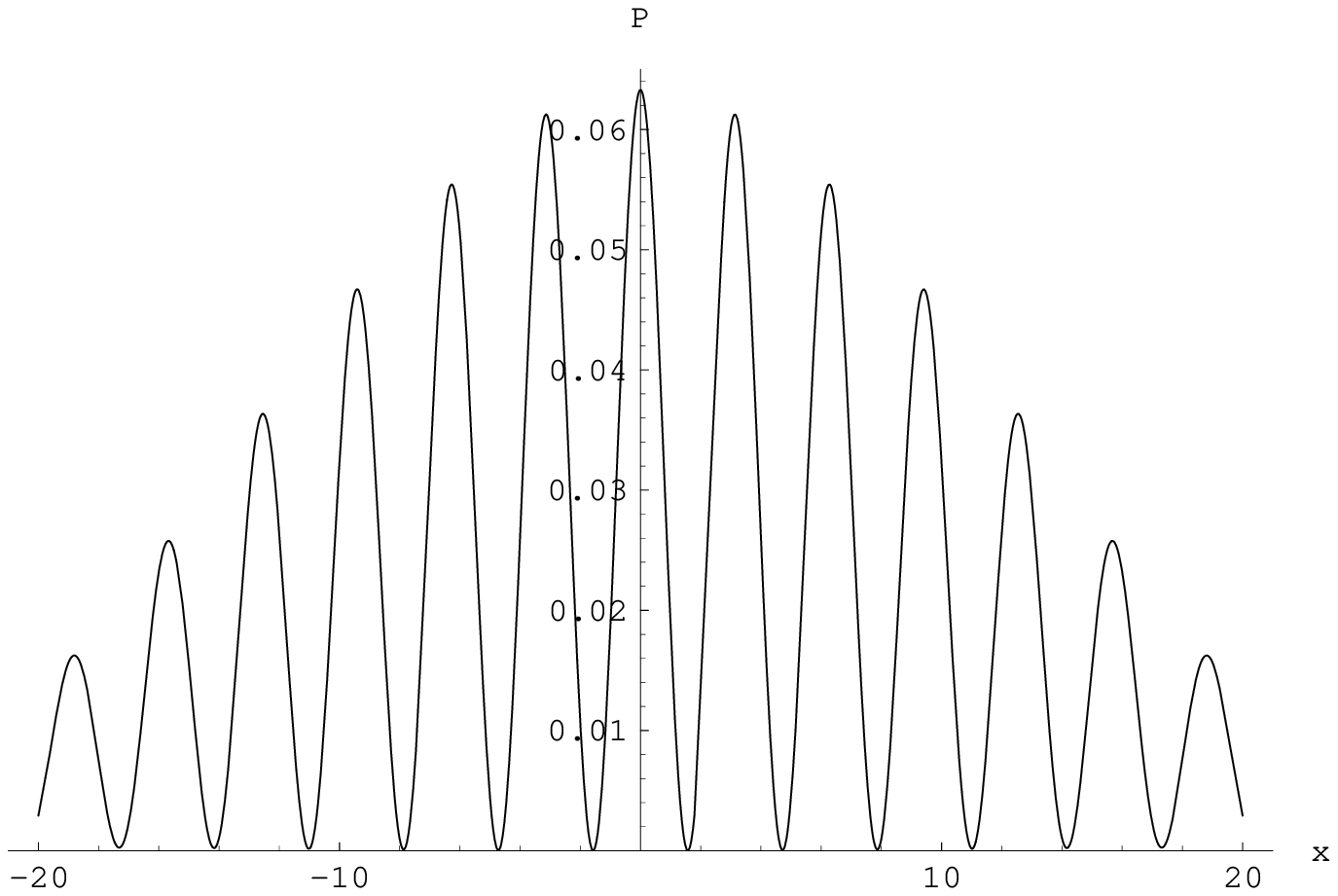}
\caption{Quantum Two-Slit Experiment} \label{quantum2}
\end{figure}
\end{center}


\begin{thebibliography}{99}
\bibitem{Wigner}
E.P. Wigner, Phys. Rev. {\bf 40}, 749 (1932);
\bibitem{Moyal}
J.E. Moyal, Proc. Camb. Phil. Soc. {\bf 45}, 99 (1947);
\bibitem{Koopman}
B.O. Koopman, Proc. Natl. Acad. Sci. U.S.A. {\bf 17}, 315 (1931);
\bibitem{von Neumann}
J. von Neumann, Ann. Math. {\bf 33}, 587 (1932); ibid.\,{\bf 33}, 789 (1932);
\bibitem{Gozzi2}
E. Gozzi, M. Reuter and W.D. Thacker, Phys. Rev. D {\bf 40}, 3363 (1989);
\bibitem{Sakurai}
J.J. Sakurai, {\it Modern Quantum Mechanics},
Addison-Wesley, Reading, MA,1995; 
\bibitem{Sudarshan}
T.N. Sherry and E.C.G. Sudarshan, Phys. Rev. D {\bf 18} , 4580 (1978);
\bibitem{Messiah}
A. Messiah, {\it Quantum Mechanics}, North-Holland, Amsterdam, 1961;
\bibitem {Gozzi3}
E. Gozzi, Private communication;
\bibitem{Abrikosov}
A.A. Abrikosov and E. Gozzi, Nucl. Phys. B Proc. Suppl. {\bf 88}, 369 (2000);\newline 
quant-ph/9912050;
\bibitem{Feynman}
R.P. Feynman, {\it The Character of Physical Law}, MIT-Press, Cambridge, 1967;
\bibitem{Hibbs}
R.P. Feynman, {\it Quantum mechanics and Path Integrals}, McGraw-Hill, New York, 1965;
\bibitem{last}
E. Gozzi and D. Mauro, work in progress.
\end{thebibliography}
\end{document}